**Author for correspondence:**
Xing Wei
e-mail: xingwei@bnu.edu.cn
ORCID: 0000-0002-3641-6732

# Energy partition in magnetohydrodynamic turbulence

## Xing Wei

IFAA, School of Physics and Astronomy, Beijing Normal University, Beijing, China

We use a simple and straightforward method to derive the energy partition in magnetohydrodynamics (MHD) turbulence that was first studied by Lee and then more rigorously by Chandrasekhar. By investigating the energy equation we find that the turbulent viscous and ohmic dissipations are comparable to each other. Under the condition that turbulent viscosity and turbulent magnetic diffusivity are comparable, we deduce that the ratio of kinetic to magnetic energies depends on the ratio of the turbulent magnetic lengthscale to turbulent velocity lengthscale of the largest eddies. When the two largest lengthscales are comparable, the two energies are in equipartition.

## 1. Introduction

Magnetohydrodynamic (MHD) turbulence (e.g., Biskamp [1]) widely exists in astrophysics (e.g., stellar convection zone, stellar wind, star formation region, accretion disk) and geophysics (e.g., earth's fluid core). How the turbulent kinetic and magnetic energies are partitioned in MHD turbulence is a classical problem dated back to TD Lee, who first derived the equipartition of two energies using the energy spectra [2]. As a famous particle and statistical physicist who passed away last year, TD Lee also made great contributions to astronomy and fluid dynamics. In astronomy, he reduced Chandrasekhar limit from 5.72 to 1.44 solar mass by considering helium-rich white dwarfs [3]. In fluid dynamics, he pointed out that 2D turbulence is quite different from 3D turbulence because of the absence of vortex stretching term [4], and that kinetic and magnetic energies in MHD turbulence are in equipartition [2]. In Lee [2] he assumed that fluid velocity and magnetic field do not transfer energy at the same lengthscale in the inertial range, so that they are in equipartition. Then Chandrasekhar [5] used structure functions to give a more rigorous derivation for homogeneous isotropic MHD turbulence, and found that the two energies in the inertial range are almost in equipartition, i.e., magnetic energy $\sim 1.63$ kinetic energy. In the next years, the energy equipartition in MHD turbulence was verified by the numerical simulations of dynamo and magneto-rotational instability [e.g., 6; 7; 8, etc.]. In the magneto-rotational simulations magnetic energy is around 5 times kinetic energy but they are still at the same order [e.g., 7]. There are also some opposite results. For example, Brandenburg [9] found that the equipartition in compressible MHD turbulence can be broken and the ratio of two energies depends on the microphysical magnetic Prandtl number, namely the ratio of laminar viscosity to magnetic diffusivity.

The energy partition in MHD turbulence is important in astrophysics, solar physics, space physics, etc. For example, how the energies are partitioned in solar-wind driven turbulence is crucial for understanding whether the corona heating is induced by Alfvén wave or magnetic reconnection or some other mechanisms, which is the major scientific purpose of Parker solar probe. In stellar dynamo, how magnetic field relates to rotation depends on the energy partition [10]. In accretion disk, energy partition can be used to judge whether accretion is induced by magneto-rotational instability or magnetic wind [7]. However, some researchers in astrophysics and geophysics are not very familiar with the turbulence theory, e.g., the structure functions or the Karmen-Howarth equation. To better understand the energy partition in MHD turbulence seems necessary for astrophysics and geophysics. Moreover, such understanding will also provide some insightful physics (see Section 3).

In this small paper, we study the incompressible MHD turbulence driven by an external force in the absence of rotation and stratification, and turbulence is already in a statistically steady state. We derive the energy equipartition in a straightforward manner and avoid the rigorous structure functions.

## 2. Formulation

We start from the momentum equation of turbulent incompressible MHD

$$\rho(\partial_t \boldsymbol{u} + \boldsymbol{u} \cdot \boldsymbol{\nabla} \boldsymbol{u}) = -\boldsymbol{\nabla} p + \rho \nu_t \nabla^2 \boldsymbol{u} + \boldsymbol{J} \times \boldsymbol{B} + \boldsymbol{f} \qquad (2.1)$$

where $\boldsymbol{J}$ is electric current, $\boldsymbol{B}$ magnetic field, $\boldsymbol{f}$ the external force to drive turbulence, e.g., buoyancy force $\boldsymbol{f} = \delta\rho\boldsymbol{g}$ for dynamo, and $\nu_t$ is turbulent viscosity. Turbulent Reynolds stress is modeled with turbulent viscosity, $-\langle u'_i u'_j \rangle = \nu_t \partial_j \langle u_i \rangle$ where $\langle \rangle$ denotes the ensemble average and prime denotes the fluctuation. Performing $\boldsymbol{u}\cdot$ on the momentum equation we obtain the kinetic energy equation

$$\partial_t(\rho u^2/2) + \boldsymbol{\nabla} \cdot ((\rho u^2/2)\boldsymbol{u}) = -\boldsymbol{\nabla} \cdot (p\boldsymbol{u}) - \rho\nu_t(\partial_j u_i)^2 + \boldsymbol{u} \cdot (\boldsymbol{J} \times \boldsymbol{B}) + \boldsymbol{f} \cdot \boldsymbol{u} \qquad (2.2)$$

where $\partial_j u_i$ in the viscous dissipation term denotes the velocity gradient and its square takes Einstein's summation rule. The work done by Lorentz force $\boldsymbol{u} \cdot (\boldsymbol{J} \times \boldsymbol{B})$ can be written as

$$\boldsymbol{u} \cdot (\boldsymbol{J} \times \boldsymbol{B}) = -\boldsymbol{J} \cdot (\boldsymbol{u} \times \boldsymbol{B}) = -\boldsymbol{J} \cdot (\boldsymbol{J}/\sigma_t - \boldsymbol{E}) = \boldsymbol{J} \cdot \boldsymbol{E} - J^2/\sigma_t \quad (2.3)$$

where $\sigma_t$ is turbulent electric conductivity and Ohm's law $\boldsymbol{J} = \sigma_t(\boldsymbol{E} + \boldsymbol{u} \times \boldsymbol{B})$ is employed. On the other hand, by performing $\boldsymbol{B}\cdot$ on Faraday's law $\partial_t \boldsymbol{B} = -\boldsymbol{\nabla} \times \boldsymbol{E}$, we obtain the magnetic energy equation

$$\partial_t(B^2/2\mu) = -\boldsymbol{B} \cdot (\boldsymbol{\nabla} \times \boldsymbol{E}/\mu) = -\boldsymbol{\nabla} \cdot (\boldsymbol{E} \times \boldsymbol{B}/\mu) - \boldsymbol{J} \cdot \boldsymbol{E} \quad (2.4)$$

where Ampere's law $\boldsymbol{\nabla} \times \boldsymbol{B} = \mu \boldsymbol{J}$ is employed. Adding the two energy equations, we are led to the total energy equation

$$\partial_t(\rho u^2/2 + B^2/2\mu) = -\boldsymbol{\nabla} \cdot \boldsymbol{A} - \rho \nu_t (\partial_j u_i)^2 - J^2/\sigma_t + \boldsymbol{f} \cdot \boldsymbol{u} \quad (2.5)$$

where $\boldsymbol{A}$ consists of kinetic energy flux $(\rho u^2/2)\boldsymbol{u}$, pressure energy flux $p\boldsymbol{u}$, and Poynting flux $\boldsymbol{E} \times \boldsymbol{B}/\mu$. The term $\boldsymbol{J} \cdot \boldsymbol{E}$ in the work done by Lorentz force is cancelled. The term $J^2/\sigma_t$ is ohmic dissipation rate.

Next, we take the volume average of the total energy equation by integrating over the whole domain of MHD turbulence, e.g., the stellar convection zone or accretion disk. At the statistically steady state, the time derivative $\partial_t$ on the left-hand-side vanishes. The fluxes $\boldsymbol{A}$ across the boundaries are tiny compared to the internal energy and thus neglected. Consequently we achieve the balance

$$\boldsymbol{f} \cdot \boldsymbol{u} \simeq \rho \nu_t (\partial_j u_i)^2 + J^2/\sigma_t \quad (2.6)$$

which states that the power of external force $\boldsymbol{f} \cdot \boldsymbol{u}$ is balanced by the turbulent viscous and ohmic dissipations at the statistically steady state.

In the standard turbulence theory [11], the turbulent energy flux is a constant independent of lengthscale and eventually dissipated by molecular viscosity at the smallest lengthscale. Thus, we can estimate the viscous dissipation rate which is comparable to the turbulent energy flux at the largest length scale,

$$\rho \nu_t (\partial_j u_i)^2 \sim \rho \nu_t (u/l_u)^2 \quad (2.7)$$

where $u$ is typical velocity and $l_u$ is the lengthscale of largest turbulent eddies. We will see in the next paragraph that dissipation rate is comparable to energy flux $\rho u^3/l_u$. We then investigate the ohmic dissipation rate. By Ampere's law $\boldsymbol{\nabla} \times \boldsymbol{B} = \mu \boldsymbol{J}$, we make the estimation $J \sim B/(\mu l_B)$. Here we introduce the lengthscale of magnetic field $l_B$. The ohmic dissipation rate is then estimated as

$$J^2/\sigma_t \sim \eta_t (B^2/\mu)/l_B^2 \quad (2.8)$$

where $\eta_t = 1/(\sigma_t \mu)$ is turbulent magnetic diffusivity. The turbulence magnetic diffusivity $\eta_t$ is associated with the correlation of turbulent velocity $u'$ and field $B'$, which is described by the mean field theory $\langle \boldsymbol{u}' \times \boldsymbol{B}' \rangle = \alpha \langle \boldsymbol{B} \rangle - \eta_t \boldsymbol{\nabla} \times \langle \boldsymbol{B} \rangle$. On the other hand, the power of external force, e.g., the thermal convection in star or the shear of Keplerian motion in disk, is usually on a large lengthscale, i.e., the mixing length $l_u$. In astrophysics, the mixing length $l_u$ is at the order of the scale height of density (say, in accretion disk [12]) or of pressure (say, in stellar convection zone [13]). Since the external energy on large lengthscale drives turbulence, its power is estimated to be the rate of turbulent energy of largest eddies

$$\boldsymbol{f} \cdot \boldsymbol{u} \sim \rho u^3/l_u. \quad (2.9)$$

Inserting the above estimations (2.7), (2.8) and (2.9) into (2.6), we are led to the balance

$$\rho u^3/l_u \simeq \nu_t \rho u^2/l_u^2 + \eta_t (B^2/\mu)/l_B^2. \quad (2.10)$$

In the standard turbulence theory [11], turbulent energy dissipates on the smallest scale (Kolmogorov scale) and dissipation is modeled with turbulent viscosity. The large eddies contain more turbulent energy and contribute more to turbulent viscosity, such that turbulent viscosity is estimated with the largest eddies, i.e., $\nu_t \sim u l_u$, and hence the left-hand-side $\rho u^3 / l_u$ is comparable to the first term $\nu_t \rho u^2 / l_u^2$ on the right-hand-side. Since the second term ohmic dissipation $\eta_t (B^2/\mu)/l_B^2$ is not negligible in MHD turbulence, to achieve the three-term balance (2.10), it can be neither much stronger nor much weaker than viscous dissipation. Consequently, the only possibility is that ohmic dissipation is comparable to viscous dissipation which is also comparable to external power

$$\rho u^3 / l_u \sim \nu_t \rho u^2 / l_u^2 \sim \eta_t (B^2/\mu)/l_B^2. \quad (2.11)$$

The second estimation $\nu_t \rho u^2 / l_u^2 \sim \eta_t (B^2/\mu)/l_B^2$ yields the ratio of magnetic to kinetic energies

$$(B^2/\mu)/(\rho u^2) \sim (\nu_t/\eta_t)(l_B/l_u)^2. \quad (2.12)$$

In the MHD turbulence, it seems plausible that turbulent magnetic diffusivity $\eta_t$ is comparable to turbulent viscosity $\nu_t$ (i.e., turbulent magnetic Prandtl number is at order of unity), because the former transporting magnetic flux and the latter transporting momentum are both caused by turbulence. The observational constraints [14] and numerical simulations [15; 16] have already shown that in a solar-like star turbulent magnetic diffusivity and turbulent viscosity are at the same order of magnitude, namely $10^{12}$ cm$^2$/s. We immediately obtain the energy ratio

$$(B^2/\mu)/(\rho u^2) \sim (l_B/l_u)^2. \quad (2.13)$$

It indicates that the ratio of two energies depends on the ratio of two lengthscales. This is a perspective to understand the solar-wind turbulence in which the eddies of magnetic field are larger than those of velocity [17] and the Alfven ratio (the ratio of velocity fluctuation to magnetic fluctuation) is smaller than 1 [18].

Usually, for largest eddies in astrophysical or geophysical situation, magnetic Reynolds number $Rm = u l_u / \eta$ (here $\eta$ is magnetic diffusivity in laminar flow) is very large, $Rm \gg 1$, such that Alfvén's frozen-in theorem works, that is, fluid elements are frozen in magnetic field lines (once two fluid elements are connected through a field line they will be connected through this field line all the time). The surface of the largest turbulent eddies can be considered as fluid elements, such that we can draw a conclusion that the two lengthscales $l_u$ and $l_B$ are comparable. Therefore, we achieve the energy equipartition in MHD turbulence.

## 3. Discussion

In this short paper we avoid the complex structure functions and equations of turbulence energy spectrum to achieve the energy equipartition in MHD turbulence by simply building some balances. Although our derivation is straightforward, it provides us some insightful physics. The first point is that the energy balance lies between external power and two turbulent dissipations (2.6). The second point is that the two turbulent dissipations are comparable (2.10). The third point is that both the turbulent dissipations can be estimated with the largest lengthscales (see the details about the estimations of the two turbulent dissipations). The fourth point is that the largest lengthscales of flow and field are comparable, which is a natural consequence of frozen-in theorem. These are what we learned from this simple derivation.

Some caveats need to be discussed. What we consider is homogeneous isotropic MHD turbulence, i.e., magnetic field is randomly distributed but does not have a preferred direction. In the anisotropic MHD turbulence, e.g., magneto-convection with an externally imposed field, the eddies are elongated along the imposed field lines due to the Alfvén wave propagation, such that the energy equipartition is complicated [e.g., 19], and we cannot use this simple derivation. Another situation we do not consider is compressible MHD turbulence, in which turbulent energy is decomposed into incompressible and compressible parts and shock may play an important role in dissipation [e.g., 20], such that this simple derivation cannot be used either. Moreover, we do not consider the influence of rotation or stratification on MHD turbulence either, which may bring more complex anisotropy. Finally, the decaying turbulence without driving force needs to be studied, which can never be in a statistically steady state but always time-dependent.

# References

1. Dieter Biskamp. *Magnetohydrodynamic Turbulence*. Cambridge University Press, 2003.
2. T. D. Lee. On some statistical properties of hydrodynamical and magneto-hydrodynamical fields. *Quarterly of Applied Mathematics*, 10:69–74, 1952.
3. T. D. Lee. Hydrogen Content and Energy-Productive Mechanism of White Dwarfs. *Astrophysical Journal*, 111:625, 1950.
4. T. D. Lee. Difference between Turbulence in a Two-Dimensional Fluid and in a Three-Dimensional Fluid. *Journal of Applied Physics*, 22:524–524, 1951.
5. S. Chandrasekhar. The partition of energy in hydromagnetic turbulence. *Annals of Physics*, 2:615–626, 1957.
6. A. Pouquet, U. Frisch, and J. Leorat. Strong MHD helical turbulence and the nonlinear dynamo effect. *Journal of Fluid Mechanics*, 77:321–354, 1976.
7. Steven A. Balbus and John F. Hawley. Instability, turbulence, and enhanced transport in accretion disks. *Reviews of Modern Physics*, 70:1–53, 1998.
8. E. T. Vishniac. The Nonlinear $\alpha$ - $\Omega$ Dynamo. In Elisabete M. de Gouveia dal Pino, Germán Lugones, and Alexander Lazarian, editors, *Magnetic Fields in the Universe: From Laboratory and Stars to Primordial Structures.*, volume 784, pages 3–15. AIP, 2005.
9. Axel Brandenburg. Magnetic Prandtl Number Dependence of the Kinetic-to-magnetic Dissipation Ratio. *Astrophysical Journal*, 791:12, 2014.
10. Xing Wei. Estimations and Scaling Laws for Stellar Magnetic Fields. *Astrophysical Journal*, page 40, 2022.
11. Lev Davidovich Landau and E. M. Lifshitz. *Fluid mechanics*. Butterworth-Heinemann, 1987.
12. N. I. Shakura and R. A. Sunyaev. Black holes in binary systems. Observational appearance. *Astronomy and Astrophysics*, 24:337–355, 1973.
13. Rudolf Kippenhahn and Alfred Weigert. *Stellar Structure and Evolution*. Springer-Verlag, 1990.
14. Jie Jiang, Piyali Chatterjee, and Arnab Rai Choudhuri. Solar activity forecast with a dynamo model. *Monthly Notice of Royal Astronomical Society*, 381:1527–1542, 2007.
15. T. A. Yousef, A. Brandenburg, and G. Rüdiger. Turbulent magnetic Prandtl number and magnetic diffusivity quenching from simulations. *Astronomy and Astrophysics*, 411:321–327, 2003.
16. P. J. Käpylä, M. Rheinhardt, A. Brandenburg, and M. J. Käpylä. Turbulent viscosity and magnetic Prandtl number from simulations of isotropically forced turbulence. *Astronomy and Astrophysics*, 636:A93, 2020.
17. Honghong Wu, Chuanyi Tu, Xin Wang, Jiansen He, and Linghua Wang. 3D Feature of Self-correlation Level Contours at $10^{10}$ cm Scale in Solar Wind Turbulence. *Astrophysical Journal*, 882:21, 2019.
18. Roberto Bruno and Vincenzo Carbone. The Solar Wind as a Turbulence Laboratory. *Living Reviews in Solar Physics*, 10:2, 2013.
19. Jungyeon Cho. *The structure of magnetohydrodynamic turbulence*. PhD thesis, University of Texas, Austin, 2000.
20. James M. Stone, Eve C. Ostriker, and Charles F. Gammie. Dissipation in Compressible Magnetohydrodynamic Turbulence. *Astrophysical Journal Letters*, 508:L99–L102, 1998.